\documentclass{iau} 
\usepackage{graphicx}

\title[JD 11.~The interplay between mass-loss and binarity] 
{The interplay between mass-loss and binarity}

\author[Hugues Sana]   
{Hugues Sana$^1$}

\affiliation{$^1$ Institute of Astrophysics, KU Leuven, Celestijnlaan 200D, 3001 Leuven, Belgium \\email: {\tt hugues.sana@kuleuven.be}}

\pubyear{2022}
\volume{xxx}  
\jname{Title of your IAU Symposium}
\editors{A.C. Editor, B.D. Editor \& C.E. Editor, eds.}
\begin{document}
\maketitle

\begin{abstract}
Most stars with birth masses larger than that of our Sun belong to binary or higher order multiple systems. Similarly, most stars have stellar winds.  Radiation pressure and multiplicity  create outflows of material that remove mass from the primary star and inject it into the interstellar medium or transfer it to a companion. Both have strong impact on the subsequent evolution of the stars, yet they are often studied separately. In this short review, I will sketch part of the landscape of the interplay between stellar winds and binarity.  I will present several  examples where binarity shapes the stellar outflows, providing new opportunities to understand and measure mass loss properties. Stellar winds spectral signatures often help clearly identifying key stages of stellar evolution. The multiplicity properties of these stages then shed a new light onto evolutionary connections between the different categories of evolved stars.
\end{abstract}

\firstsection 
\section{Introduction}


Stars with birth masses larger than about twice the mass of our Sun (spectral types OBA on the main sequence) are fundamental cosmic engines. These intermediate- and high-mass stars heat and enrich the interstellar medium, they drive the chemical evolution of galaxies and, at the high-mass end, their end-of-life explosions are bright enough that they can be seen  throughout cosmic times.
Given their decisive role in a wide range of astrophysical phenomena, it is of paramount importance to understand how these stars live and die.  

The lifetimes of intermediate and massive stars spans three orders of magnitudes: from about 2~Gyr for a 2~M$_\odot$ star, to just over 2~Myr for stars with inital masses of 100~M$_\odot$ or more. While vastly different, these times scales are significantly shorter than the age of the universe. As a consequence, their full evolutionary cycle, from their formation in dense molecular clouds to their final stages where (most of) these stars  shed their envelopes through strong stellar winds, can be directly observed, providing us with direct insight into their evolution.  Yet, many uncertainties remain that affect predictions of stellar evolution models. Among these, the issues related to their internal structure, rotation and mixing properties, the strengths and structure of their outflows and their stellar multiplicity are some of the most problematic, preventing proper astrophysical understanding of stellar evolution.

In this paper, we start from the postulate that stellar outflows and multiplicity are not only ubiquitous phenomena among intermediate and high-mass stars, but that much is to be gained to consider them as just but two sides of the same coin: that of the physics of the objects that host them. This paper is organised as follows.  In Sect.~\ref{s:cwb},  I will present several  examples where binarity shapes the stellar outflows, providing new opportunities to understand and measure mass loss properties. In Sect.~\ref{s:evol}, we will use the spectral signatures resulting from stellar outflows to help identifying key stages of stellar evolution. The multiplicity properties at each of these stages then shed a new light onto stellar evolution, possibly raising new questions on the nature and roles of different categories of evolved stars.

\section{Stellar outflows shaped by binarity}\label{s:cwb}

In binary systems where each of the stars in the system has a strong stellar wind,  typically O+O and WR+O binaries,
the winds from the two stars   collide at supersonic velocities, creating a wind collision region the shape of which is defined by the ram pressure equilibrium  surface between the two winds (\cite{U92,SBP92}). For WR+O system, a simplified but useful 3D geometrical model of such collision zone is provided by the cone-shock model (\cite{L97}) where the star with the weaker wind is carving a cavity in the wind of the primary star.  Both theoretical considerations (\cite[e.g., Usov (1992)]{U92})  and hydrodynamical simulations (e.g., \cite{P09,PG11}) however reveal  that the reality  is more complex: there are  two shock fronts, one for the collision of each wind with the wind-interaction region, i.e., the denser and hotter downstream region in which  the gas has been compressed and heated up by the aforementioned shocks. These simulations further reveal the presence of various instabilities (shear,  Rayleigh-Taylor, Kelvin-Helmholtz, thin-shell, ...) that can dramatically modify the appearance and the properties of the collision (\cite{P09,MGO13}), from a relatively stable configuration in the adiabatic case (e.g., WR 22, \cite{PG11}, HD93403, \cite{RVS02} to a chaotic wind collision structure in the isothermal case (e.g., HD152248, \cite{SSG04}). The main parameter delineating these two extremes depends on the speed at which the shocked gas cools down with respect to how fast it escapes the collision zone (\cite{SBP92}). This in turn strongly depends on the separation between the two stars, hence on the orbital period ($P$). In highly eccentric systems, the wind collision region can actually transition for an adiabatic to a radiative  collision zone as the two stars near the periastron passage (\cite{LDL12}, \cite{CMK15}, \cite{CMK15a}).

Typical observational signatures of these wind-wind collision include thermal X-ray emission, as the shocked gas is heated up to temperatures of 10 million Kelvin, as well as synchrotron (in the radio domain) or inverse Compton (in the hard X-rays domain) emission resulting from the presence of populations of electrons that have been accelerated to relativistic speed in the vicinity of the shock fronts. The shocked regions have also been proposed as the locus of acceleration of cosmic rays. If the cooling in the collision region is quick enough, the ionised gas might recombine, creating measurable emission in the optical, infrared and UV recombination lines. In the following, I will outline a small selection of recent results illustrating the vast diversity of behaviour and observational signatures resulting from wind-wind collisions. The focus will be on colliding wind systems hosting a Wolf-Rayet star.

\begin{figure}[t!]
\begin{center}
 \includegraphics[width=5.4in]{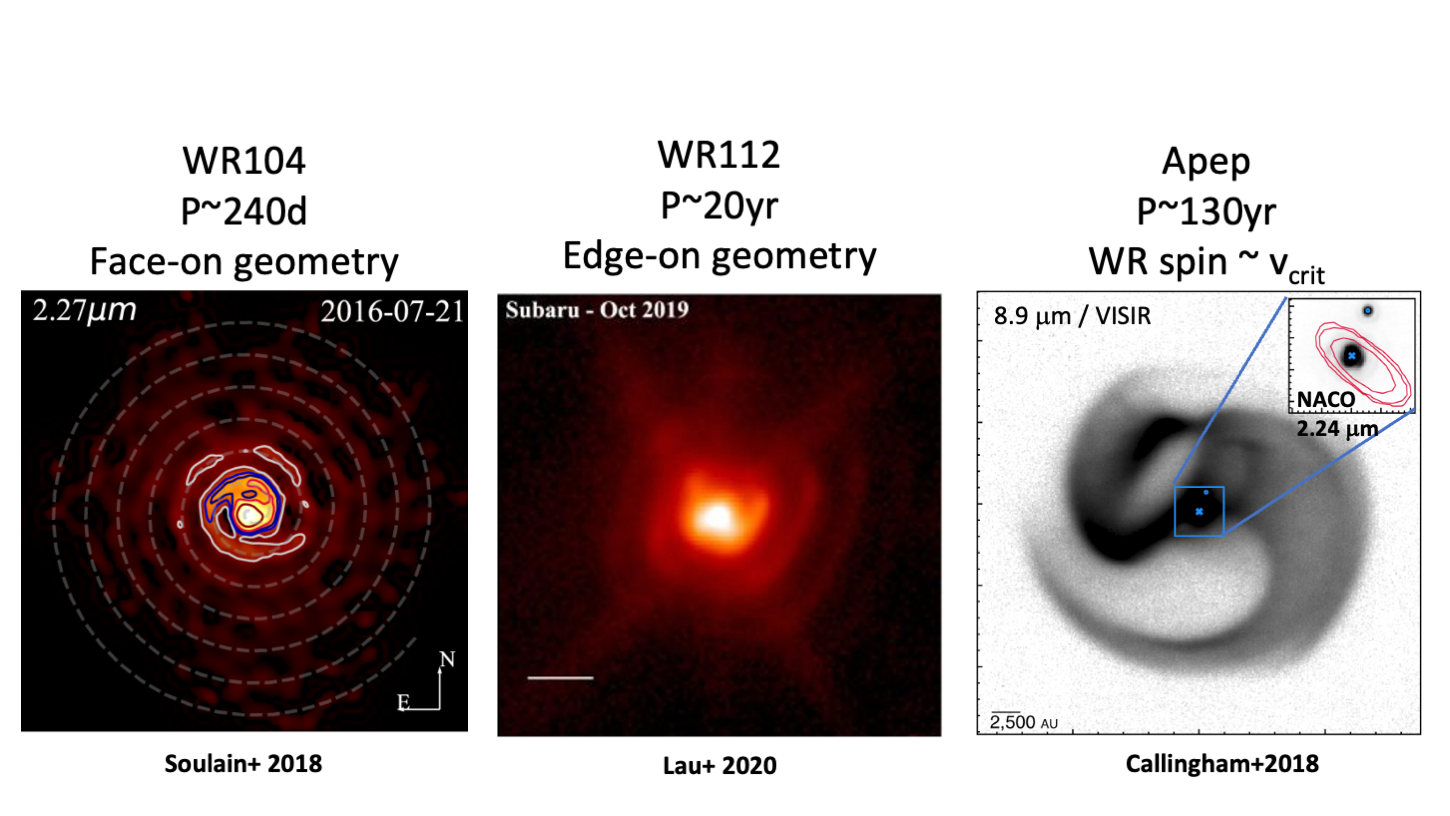} 
 \caption{ High-angular resolution images of the three WC dust producers WR~104, WR~112 and Apep, illustrating the wide diversity of morphologies and orbital periods. Figure adapted from \cite{SML18} (left), \cite{LHH20} (middle) and \cite{CTP19} (right).}
   \label{f:WCdust}
\end{center}
\end{figure}

\subsection{Episodic dust producers}

In colliding wind systems harbouring a carbon-rich Wolf-Rayet star (WC) and an O star, the enhanced density in the wind-wind collision produces favourable conditions for the nucleation of dust. Combined with the Coriolis force due to the orbital motion, this  gives rise to  so-called pinwheel nebulae which are characterised by a dust plume following an Archimedian spiral pattern.
Because the  conditions are more easily reached at periastron passage, the dust production might be episodic only. 

While the pinwheel nebulae have been know for over two decades and have been directly imaged by e.g., \cite{TMD99}, some recent works are shedding more light to  their large range of properties (see also Fig.~\ref{f:WCdust}):
\begin{enumerate}
\item[$\bullet$] WR~104 ($P \sim 241$d, \cite{TML08}) is the prototypical pinwheel nebula where the spiral structure is clearly visible thanks to the face-on orientation of the system with respect to our line of sight. First imaged in the near-infrared thanks to sparse aperture masking (\cite{TMD99}), the system has recently been observed with VLT/SPHERE and VISIR (\cite{SML18}) providing a more complete view of its spiral structure produced over the last 20~years, constraining the dust to mass ratio in the spiral arms in the range of 1 to 10\%\  and confirming that WR~104 is a hierarchical triple with the third companion  at a projected distance of about 2600~au.
\item[$\bullet$]WR ~140 is an eccentric WC7+O5 system on an $P \sim 8~$yr orbit (\cite{TRE21a}), and an episodic dust maker. It is one of the best characterised colliding wind binaries, with an almost continuously X-ray light curve illustrating the clear $1/D$ dependance of the X-ray thermal emission, with $D$ the separation between the two stars. In the radio domain, time-resolved images of the synchrotron emission from the wind interaction region shows how its position and shape change as the O star plunge into the radiosphere of the WC star while nearing periastron  (\cite{DBC05}). 
\item[$\bullet$] WR~112 ($P \sim 20$~yr) offers very different though no less spectacular images than WR ~104.  This results from its edge-on geometry. \cite{LHH20} recently constrained the dust mass-loss rate of the WC star to be about $3 \times 10^{-6}$~M$_\odot$~yr$^{-1}$, making it one of the largest dust-producing pinwheel nebula known. 
\item[$\bullet$] Apep ($P\sim130$~yr) has a very peculiar plume. Recent work by \cite{CTP19} indicates that the plume can only be reproduced by assuming that the WC object is (near-)critically rotating, making Apep a candidate gamma-ray-burst  progenitor.
\end{enumerate}

These few examples not only illustrate the large diversity of orbital periods involved (over two orders of magnitudes) but also how the structure and shape of the wind-wind collision allow astronomers to constrain the mass-loss rates of evolved massive stars (a key ingredient in stellar evolution) as well as some of the stellar properties that are needed to properly  predict their end-of-life scenarios.

\begin{figure}[t!]
\begin{center}
 \includegraphics[width=5.4in]{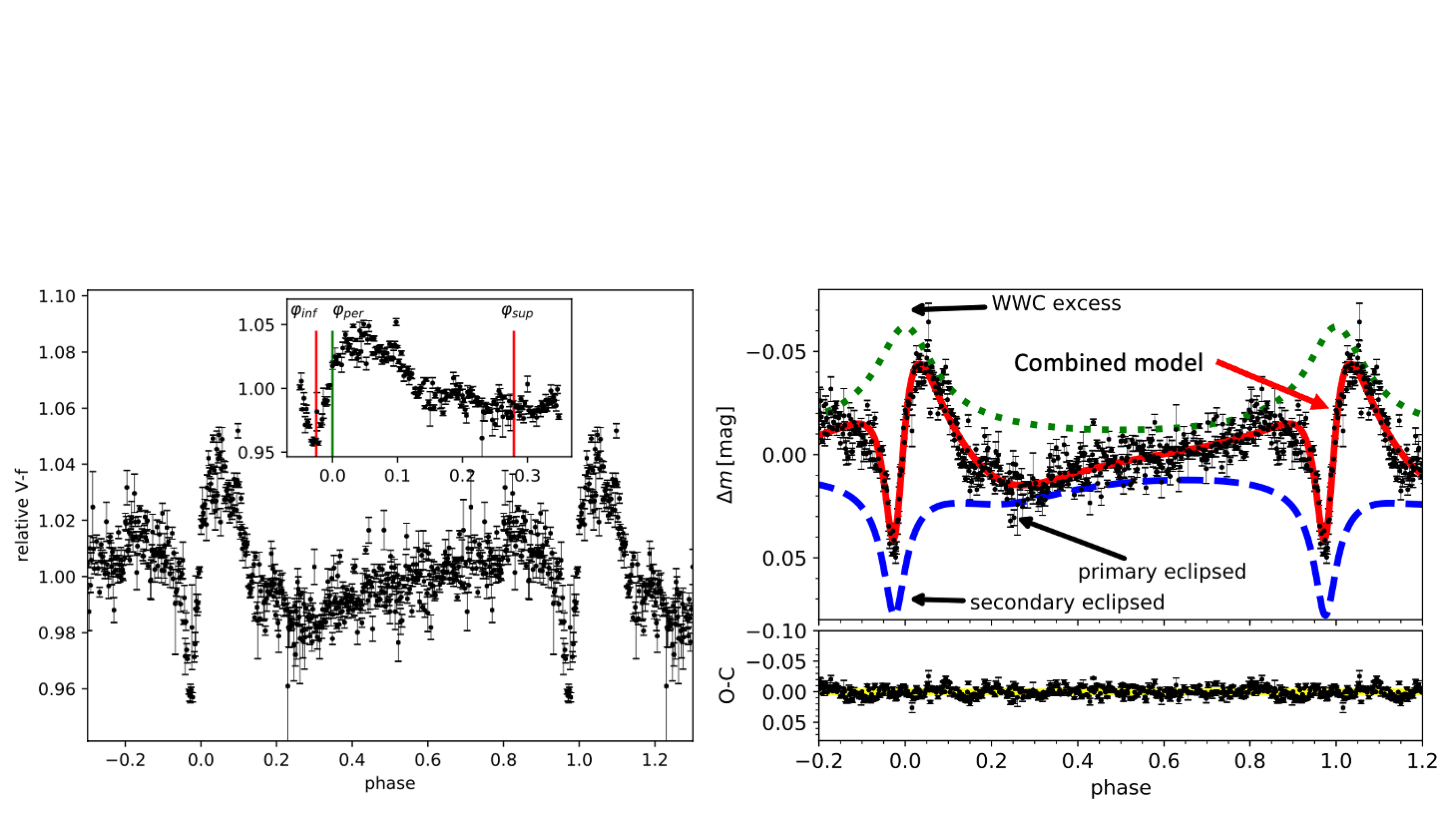} 
 \caption{{\it Left panel:} OGLE light curve of the R144 WNh5+WNh6 binary system revealing a clear heartbeat-like signature. The insert indicates the position of conjunctions ($\varphi_\mathrm{inf}$ and $\varphi_\mathrm{sup}$) and of periastron ($\varphi_\mathrm{per}$). Figure adapted from \cite{SSM21}.}
   \label{f:R144}
\end{center}
\end{figure}

\subsection{A heartbeat light curve produced by wind-wind collision}

While wind-wind collision (WWC) is known to contribute to the optical emission lines, whether in O+OB or in WR+O binary systems, its direct impact on the optical light-curve of a system is usually rather limited. Here we highlight new results from the R144 WR+WR binary in the Large Magellanic Cloud showing that this is not necessarily the case (\cite{SSM21}). 

WR~144 is a hydrogen rich Wolf-Rayet object in the Tarantula nebula. It is also the brightest X-ray source in the region and, given its predicted bolometric luminosity, has been suspected for a long time to be a binary system. One  had to wait the advent of the X-Shooter spectrograph at ESO/VLT to detect a clear binary signature, with significant radial velocity (RV) variations over time scales of months and the presence of two hydrogen rich WNh stars (\cite{SvT13}), suggesting that WR~144 is one of the most massive binary system known so far. 

Using additional data from the  Tarantula Massive Binary Monitoring program (\cite{AST17a}) from the FLAMES and XSHOOTER instruments at the VLT, \cite{SSM21} obtained the first orbital solution, with $P\sim74$~d and a significant eccentricity of $e\approx0.5$. Spectral disentangling allowed for a more precise spectral typing and atmosphere modelling, revealing a WNh5 and WNh6 components and measured mass-loss rates of the order of $4\times 10^{-5}$~M$_\odot$~yr$^{-1}$.

Given the properties of the system. it was clear that wind-wind collision was to be expected, and indeed, strong phase-locked variation of  He and N lines was observed. More intriguing though, the OGLE light curve of R144 displayed the characteristic of heartbeat systems where a strong dip in the light curve over a very short period of time is followed by a sharp rise before returning to its average level (Fig.~\ref{f:R144}). Heartbeat systems are usually interpreted as tidal deformation near periastron passage in eccentric binaries and are typically lower mass stars (\cite{TEM12}). 

A first attempt to model the heartbeat light curve with tidal deformation provided a good fit but required stellar radii that were about 10 times larger than expected for WNh stars. The second attempt turned out more successful. Indeed \cite{SSM21} showed that the light curve could be modelled by a combination of double wind eclipses and a phase-locked contribution of the order of a few tenths of magnitude from the emission lines to the optical magnitude (Fig.~\ref{f:R144}). The obtained model is very sensitive to inclination and the best fit yields $i=60\pm2^\mathrm{o}$ and $\dot{M} \approx 4$ to $5\times 10^{-5}$~M$_\odot$~yr$^{-1}$ (in excellent agreement with estimates from the spectroscopic analysis), resulting in absolute (dynamical) masses of 74 and 69 ($\pm4$) M$_\odot$ for the WNh stars in R144. This makes R144 one of the most massive binaries known and one of the few for which direct mass measurements can be obtained thanks to a careful analysis of its wind-wind collision.

\subsection{A unified scheme to explain the geometry of the outflows of AGB stars}

Asymptotic giant branch stars (AGBs) are  post-core-He-burning low and intermediate mass stars that are characterised by strong, chemically enriched winds and  extended circumstellar enveloppes. Their strong winds strip the AGB stars of their enveloppes, leading them to the post-AGB and planetary nebula (PN) phases.  In the latter phase, the hot PN central star emits copious amount of UV radiation that ionizes the ejecta, giving rise to beautiful planetary nebulae. Aside from their certain aesthetical qualities, one of the fascinating aspects of these planetary nebulae is the wide diversity of shapes encountered.

Similarly spectacular, ALMA observations are now able to reveal similar diversities of morphologies around AGB stars (\cite{DMR20}), suggesting that a common physical mechanism is  shaping the outflows of both AGBs ad PNe. The mass-loss rates of the AGB outflows can be well measured with modern ALMA data. Such mass-loss rates provide a reference clock with which, one can show, the AGB morphologies are correlated (\cite{DMR20}). This key result suggests that there is an evolutionary connection between the different morphologies of AGB winds. A ``young'' AGB in a binary system will likely produce an equatorial density enhancement. As the system ages and the mass loss increases, the orbit will widen due to angular momentum loss leading first to a bipolar nebula, then a spiral-like structure.  

This novel results is yet another example of the evolutionary diagnostic power that can be obtained by combining stellar outflow and orbital properties. We will come back to another such case for higher mass stars in Sect. \ref{s:evol}.

\subsection{Wind-wind-collision and the habitability of Earth-like planets}

While this review focuses on intermediate and high-mass stars, the following example shows that wind-wind collision is actually of broader relevance.  \cite{JZP15} studied the impact of the wind-wind collision, between the winds of two solar-type stars separated by 0.5 au, on the habitability of a circumbinary earth-like planet. The authors show that, at radial distances of about 1 to 2 au, the wind-wind collision region has density and temperature enhancements of about a factor 3 and 5, respectively, compared to the material outside the wind-wind region. A circumbinary planet orbiting the system will be swept by the collision region multiple times during its orbit, so that it spends about a quarter of its orbital period within the density- and temperature-enhanced wind-wind collision region. The main effect is to compress the magnetosphere by about 20\%. While this effect in principle remains limited for a fully formed earth-like planet, a ``young'' earth with a  denser and more expanded atmosphere might actually be partially stripped by these effects.

\section{Evolutionary connections in the upper HRD} \label{s:evol}

In this second part of this paper, we will focus on evolutionary connections between various populations of evolved stars. We will focus exclusively on massive stars, that is stars massive enough to become red supergiants and to end their life in core-collapse events. These evolved populations of massive stars are mostly identified through atmospheric  and wind-related diagnostics. Specifically, we will consider five broad categories: OB stars, red supergiants (RSGs), classical Wolf-Rayet stars (WR), luminous blue variables (LBVs) and Be stars.

In a single-star context,  the evolution of massive stars is generally described by the Conti scenario (\cite{C76a}),  which makes evolutionary connections between different categories of massive stars. A simplified version of it is given below. A more detailed version of it can be found in, e.g., \cite{LL17}.  \\

{\centering 
$40  M_\odot \lesssim  M_\mathrm{ini} \lesssim 90  M_\odot: $

$\mathrm{early\hspace*{1mm}O/Of/WNh} \rightarrow LBV \rightarrow WN \rightarrow WC (\rightarrow WO),  $ \vspace*{3mm}

$25 M_\odot  \lesssim M_\mathrm{ini} \lesssim 40 M_\odot:$

$\mathrm{mid\hspace*{1mm}O} \rightarrow YSG  \rightarrow RSG \rightarrow WN \rightarrow WC, $ \vspace*{3mm}

$8 M_\odot \lesssim  M_\mathrm{ini} \lesssim 25 M_\odot:$

$\mathrm{late\hspace*{1mm}O/early\hspace*{1mm}B} \rightarrow YSG \rightarrow RSG ( \rightarrow blue \hspace*{1mm} loop/YSG? \rightarrow  RSG).  $ \vspace*{3mm}

}

The main ingredient that defines a given evolutionary pathway is the star's initial mass. The amount of mass lost during the evolved phases is also important as it controls whether a post-hydrogen burning objects (RSG, LBV) will be stripped of its envelope, hence returning to the hotter part of the Hertszprung-Russell diagram (HRD).

While the Conti scenario looks at the connections between different evolutionary stages through a single-star, wind-focused angle, one can repeat the same exercise from different perspectives. For example, \cite{ST15}, \cite{HWD16} and \cite{S19} used the relative isolation of stars in each category. Here, we propose to use the multiplicity properties of these different categories of evolved massive stars to investigate their evolutionary connections.

\subsection{main-sequence OB stars}

In this review, we will use the ``OB star'' denomination to broadly designate hot main-sequence massive stars (\cite{WSS14}), i.e., hydrogen-burning stars with an initial mass larger than about  8 M$_\odot$. This category thus encompasses most O-type stars, B supergiants, and  B dwarfs with spectral type earlier than B3 on the zero-age main-sequence. Without loss of generality, we will include the very WNh massive stars in this category despite the fact that their spectral type is actually that of an hydrogen-rich Wolf-Rayet (WNh, \cite{dHH97}). From an evolutionarily perspective though, these WNh stars are indeed very massive hydrogen-burning stars and thus fit our definition of ``OB stars".

In the Milky Way, the multiplicity fraction of stars is know to increase with stellar mass (\cite{DK13}, \cite{MD17}), from slightly less that 50\%\ for solar type stars, to $>$60\%\ for early B-stars (\cite{AGL90, DDS15a, BSM22}) and $> 70$\%\ for O-type stars (\cite{Sdd12}). These lower limits on the multiplicity fractions increase when considering wider binaries, observed e.g. through high-angular resolution imaging. For example, \cite{SLL14} derived 90$\%$ for all O-type stars and 100\%\ for O dwarfs, a result confirmed by more recent work on the Orion and M17 star forming region by \cite{GPR13}, \cite{KPE18} and \cite{BFS22}, respectively.

To decide whether a binary system will interact and when this interaction will happen (e.g., within or beyond the main-sequence phase), the most important  property is probably  its orbital period. Of interest, the derived period distributions of O and early-B stars derived from a variety of environments are strikingly similar (\cite{SKM13,VTE21,BSM22}), pointing towards a common, metallicity independent mechanism responsible to the pairing of massive binaries. 

While these period distributions will be used as initial conditions in the following, a word of caution is needed. It is indeed possible that these distributions do not directly result from the star formation process itself. They may have been altered by early dynamical processing, through e.g.,  migration as suggested by recent results on the dearth of short-period systems in M17, one of the youngest massive star forming regions studied in depth (\cite{RKK17,SRK17}). The early dynamical processing of the orbital properties  is further supported by an observed correlation of the dynamical dispersion as a function of cluster age as put forward by \cite{RBd21}. The observed orbital  distributions of OB stars might  not be the exact end product of star formation; yet, they  probably provide good enough starting points to compute their subsequent evolution and future fate. One major uncertainty, perhaps, remain the role of triple and higher order multiples and whether dynamical effects might impact the frequency, timing, and nature of binary interactions.

\subsection{Red supergiants}

Red Supergiants (RSGs) are typically cooler than 5000~K and characterised by a strong wind outflow, the launching mechanism of which remains poorly understood and may require a combination of various processes, such as pulsation, magnetic pressure and radiation pressure on molecular lines (e.g., \cite{KSD21}). Red supergiants represent the He-burning phase of stars with initial masses typically between 8 and 25~M$_\odot$. RSGs may directly collapse and explode as supernovae or loose their envelope and turn into classical Wolf-Rayet stars.

The RSGs multiplicity is a much more difficult affair than that of their unevolved counterparts. This results not only from their fewer numbers, making statistically meaningful results difficult to reach, but also because their much larger radii. As a consequence, the orbital periods are the order of years to decades, so that corresponding radial velocity variations become of a similar magnitude as the stochastic noise expected from the large convection cells at the surface of these stars. 

\begin{figure}[t!]
\begin{center}
 \includegraphics[width=6.4in]{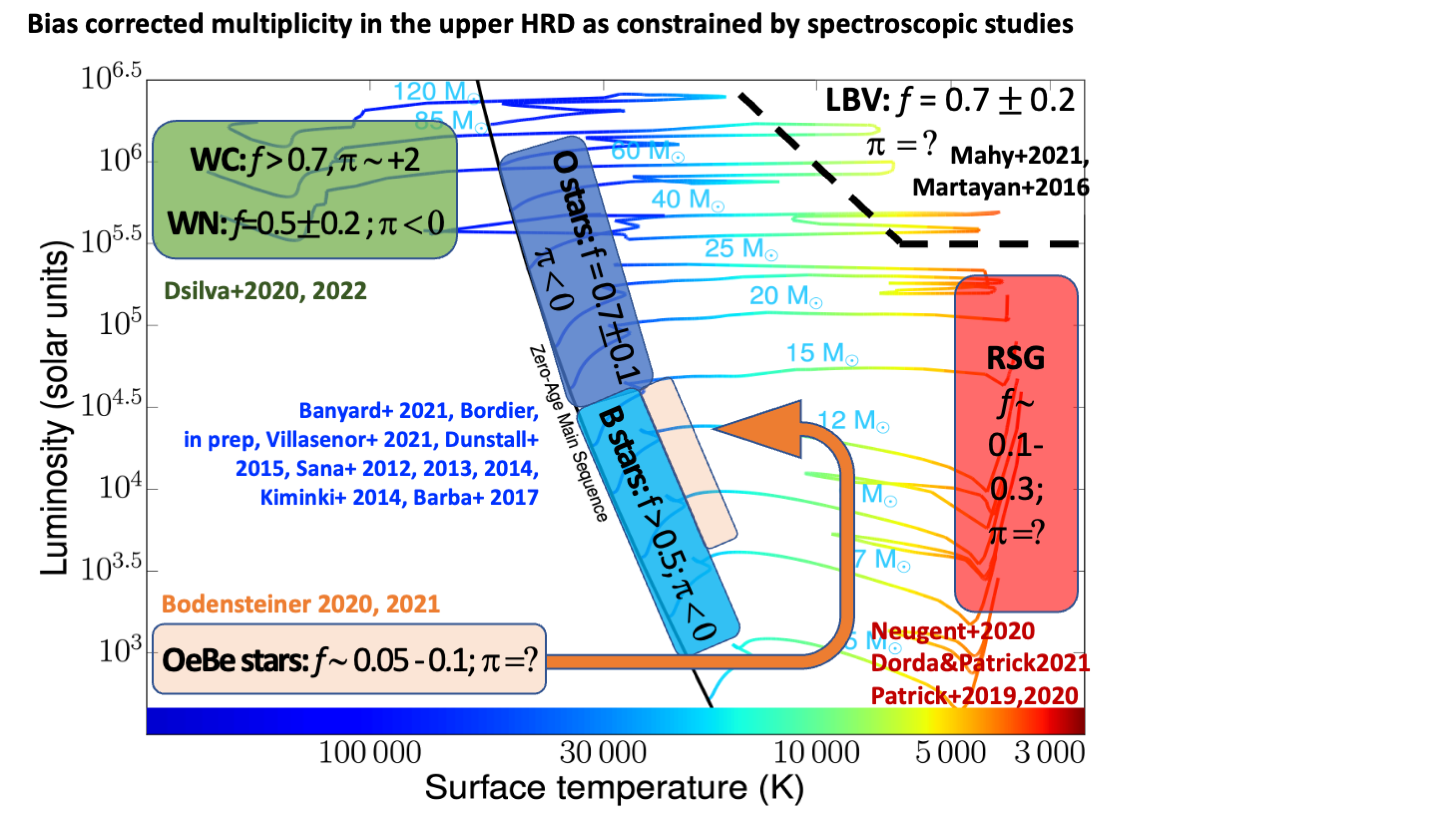} 
 \caption{Overview of our current knowledge of the multiplicity in the upper Hertszprung-Russell diagram (HRD) based on spectroscopic results. $f$ is the multiplicity fraction, $\pi$ is the index of the period distribution ($f_{\log P}\propto(\log P)^\pi$). The multiplicity of OB stars nears $f=1$ once the interferometric data are included. \cite{BdC11a} evolutionary tracks have been overlaid.}
   \label{f:HRD}
\end{center}
\end{figure}

Figure~\ref{f:HRD} summarises some of the recent results, mostly obtained in the Magellanic Clouds. These studies consistently obtained observed multiplicity fractions of the order of 10 to 20\%, and bias-corrected fractions of 20 to 30\%\ (e.g., \cite{PLB19a, PLE20a, NLM20, DP21}). These fractions are significant smaller than the $>60$ to 70\%\ claimed for their progenitors, the main-sequence OB stars with initial mass lower than $\sim25-40$~M$_\odot$. 

The lower  RSG binary fraction can be understood, at least qualitatively, by invoking binary interaction. Indeed, most binaries with orbital separation smaller than the typical size of a RSG will actually interact either during the main sequence (for $P<6$ d, approx.) or when crossing the Hertzsprung gap ($P < 2\,000$~d, approx). A simple integration of the (bias-corrected) initial period distributions presented in Fig. \ref{f:PdistWR} suggest that about 2/3 of the systems will interact. Depending on the mass ratio and nature of the interaction, we would face either a merging (resulting in a single star), a stable mass-transfer (stripping the primary of (most of) its envelope, thus   preventing the primary star to reach the RSG phase) or a common envelope ejection (also resulting in a non-RSG product). The net result is to effectively remove the close binary systems from the RSG sample, so that the multiplicity fraction of RSGs is indeed expected to be a factor 3 to 5 lower than that of their main-sequence counterparts.

While this exercise shows that there is no fundamental tension between the multiplicity fraction of OB main-sequence stars and RSGs, there is however a strong corollary: a significant number of post-interaction products produced by the interaction of binaries born as OB systems. The number of post-interaction systems known, while growing, remains consuspiciously small.
 In the next section we consider  the proposition that classical Be stars are an important  population of interaction products.

\subsection{Classical Be stars}

Be stars are B-type stars with hydrogen line emission in their spectra. While there are various types of Be stars, we will focus here on classical Be stars, i.e., Be stars whose emission is believed to be associated with a decretion ``disk". The presence of a disk is expected to result from an equatorial density enhancement  in (near)critical, rapidly rotating stars where the centrifugal forces at the equator become equal or larger than the gravitational attraction. 
There are three main scenarios to explain the origin of the classical Be phenomenon, all linked to how Be stars acquire their  (near-)critical spins: star formation (high spin at birth), single-star evolution (angular momentum transfer from the core to the envelope when the star nears the terminal age main sequence (TAMS)), angular momentum accretion due to mass-transfer in a  binary system. In the following, we will only consider the latter two.

One of the expected consequences of  stable Roche lobe overflow is the transfer of angular momentum from the initially more massive donor star to the initially least massive accreting star.  It has been shown that even a small accretion efficiency would quickly spin up the accretor to critical rotation velocities. Candidate post-interaction systems could thus host a main-sequence rapidly rotating star and classical Be stars offer a naturally match to this description. \\

Let us now investigate the predictions for the multiplicity of Be stars depending on the formation mechanism:\smallskip

{\it Scenario 1 - Binary Star Channel: } producing a Be star by mass transfer and accretion will strip the primary star from (most of) its envelope, resulting in a so-called stripped-star, i.e. an (almost) naked He core. In general, these are not very luminous and  relatively low mass objects, so they do not induce large RV variation of their Be companion. Besides, obtaining accurate RV measurements of Be stars  is notoriously difficult so that chances are that Be+He systems would likely remain undetected and appear as presumably single Be stars in RV surveys. \smallskip

{\it Scenario 2 - Single Star Channel: } At least half the early B-type stars are found in B+BA spectroscopic binary systems. If the Be star phenomenon is produced by angular momentum transfer from core to envelope when the star is nearing the TAMS, then this is expected to be a widespread phenomenon affecting all B-type binaries with period larger than about 10 days. The latter requirement on the orbital period ensures that the orbital separation is  large enough so that no interaction happens  during the main-sequence lifetime of the primary. This is indeed needed so that the B-type primary has the opportunity to become a Be star via the single-star channel before any interaction occurs. Within periods ranging from  10~days to a few 100s of days, these Be+BA binaries would appear as spectroscopic binaries that are reasonably easy to spot either through double line profiles or through large RV variations of the Be star (orbital velocity $v_\mathrm{orb}>20$~km s$^{-1}$).\smallskip

Qualitatively, the  predictions of this discussion is counter-intuitive: in the single star channel for the formation of Be stars, one expects a fair fraction of them ($>25$\%\ given some bias estimates, \cite{BSS20}) of them to be detected as spectroscopic binaries with a main-sequence, unevolved lower mass companion. Alternatively, if the binary channel is responsible for the formation of Be stars, one should observe Be stars as single stars or in binaries with an evolved (stripped) companion.

In that context, \cite{BSS20} performed a literature study of about 300 Be stars with spectral-type earlier than B1.5 and periods shorter than 5000~days. They noted that 91\%\ of the sample is reported to be single, 5\%\ have an evolved companion and 4\%\ have a companion of an unidentified nature, i.e. they could  be evolved companions or  main-sequence stars. Even if all these companions are main-sequence companions, this is significantly lower than the 25\%\ expected from the initial multiplicity properties of B-type stars and strongly suggests that the binary channel is a dominant channel to produce classical Be stars.

\begin{figure}[t!]
\begin{center}
 \includegraphics[width=5.4in]{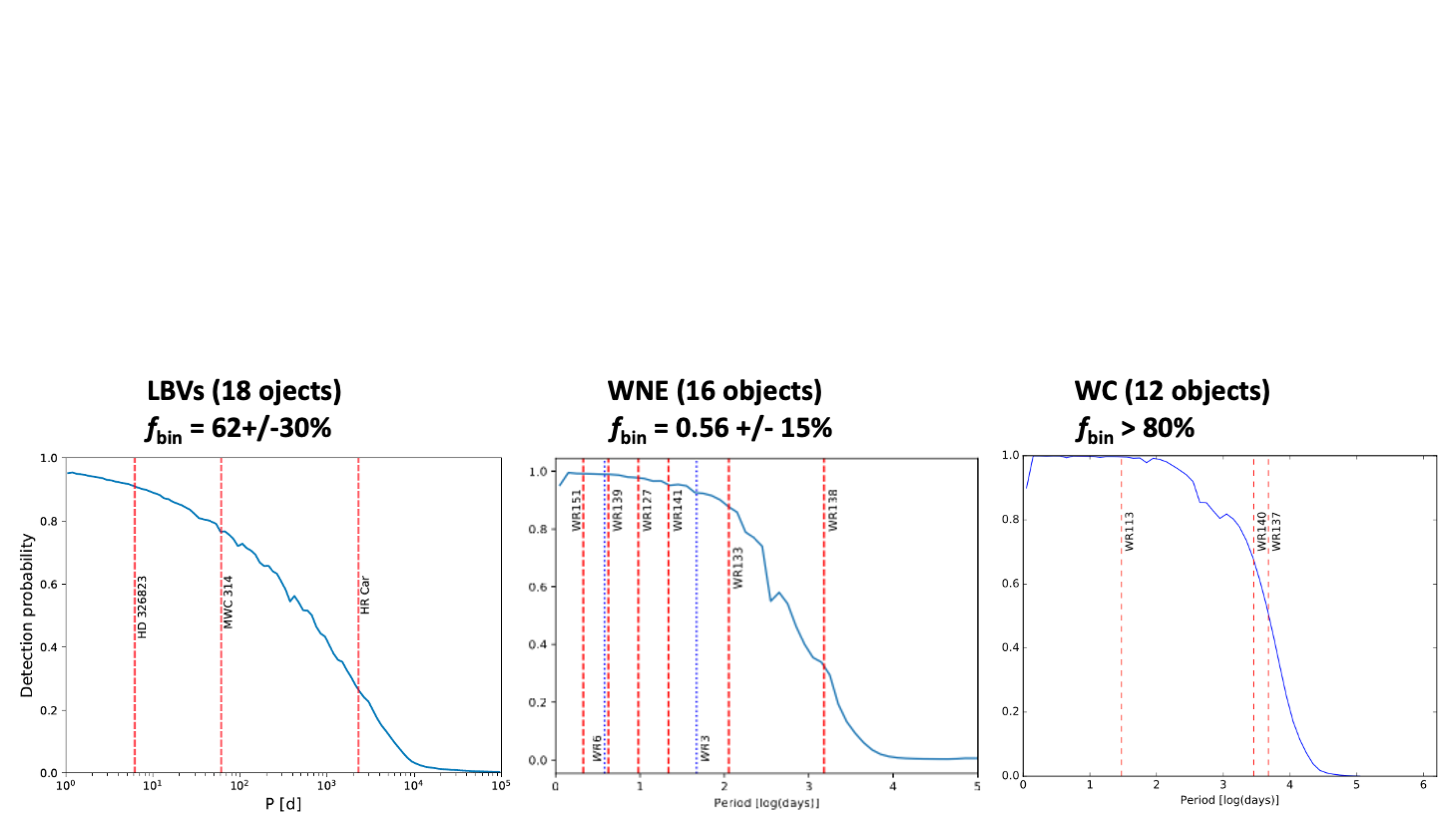} 
 \caption{Detection probability curves of the spectroscopic campaigns of \cite{MLH22} (LBV, left), \cite{DSS22} (WNE, center) and \cite{DSS20} (WC, right). The location of the objects with known orbital periods is indicated by dashed lines. Sample sizes and bias-corrected binary fractions are indicated on top of their respective plot. }
   \label{f:Pdetect}
\end{center}
\end{figure}

\subsection{Luminous Blue Variables}

 Luminous Blue Variables (LBVs) are evolved massive stars that are observationally described by their  name: luminous, blue  and variable. They are found in the upper right part of the HRD. The exact nature of LBVs and their evolutionary connection with other categories of stars remain debated. In the  main stream view, they are believed to be the evolved counterpart of very massive stars that have crossed the Hertzsprung gap after the end of hydrogen burning. By doing so, they have encountered the Humphreys-Davidson limit, a limit in the HRD where the radiative pressure at the surface becomes of the same order of magnitude as the gravitational attraction. This situation is expected to be unstable, hence the expected variabilities, and possibly leads to massive outbursts.

Using interferometry of a sample of LBV and LBV candidates, \cite{MLH22} constrained the LBV multiplicity fractions of about 60\%\ to 80\%, a value compatible with that derived from spectroscopy alone (Fig.~\ref{f:Pdetect}). Such a high binary fraction disfavours a scenario in which LBVs are predominantly runaway stars that have gained mass in a previous binary interaction. \cite{MLH22} thus suggest that LBV should  come from one of the three following channels: (i)
 single-star evolution in wide binary, i.e. wide enough to not interact; (ii)
 the result of a merger in an initially triple system where one does not detect the outer companion, or
(iii) the descendants of short-period binaries that have widen a lot due to non-conservative mass transfer.
While the multiplicity fraction of LBVs is similar to that of OB stars, one has to note that most of the detected companions around LBVs are on much wider orbits given the larger radii of LBV stars.

\begin{figure}[t!]
\begin{center}
 \includegraphics[width=6.4in]{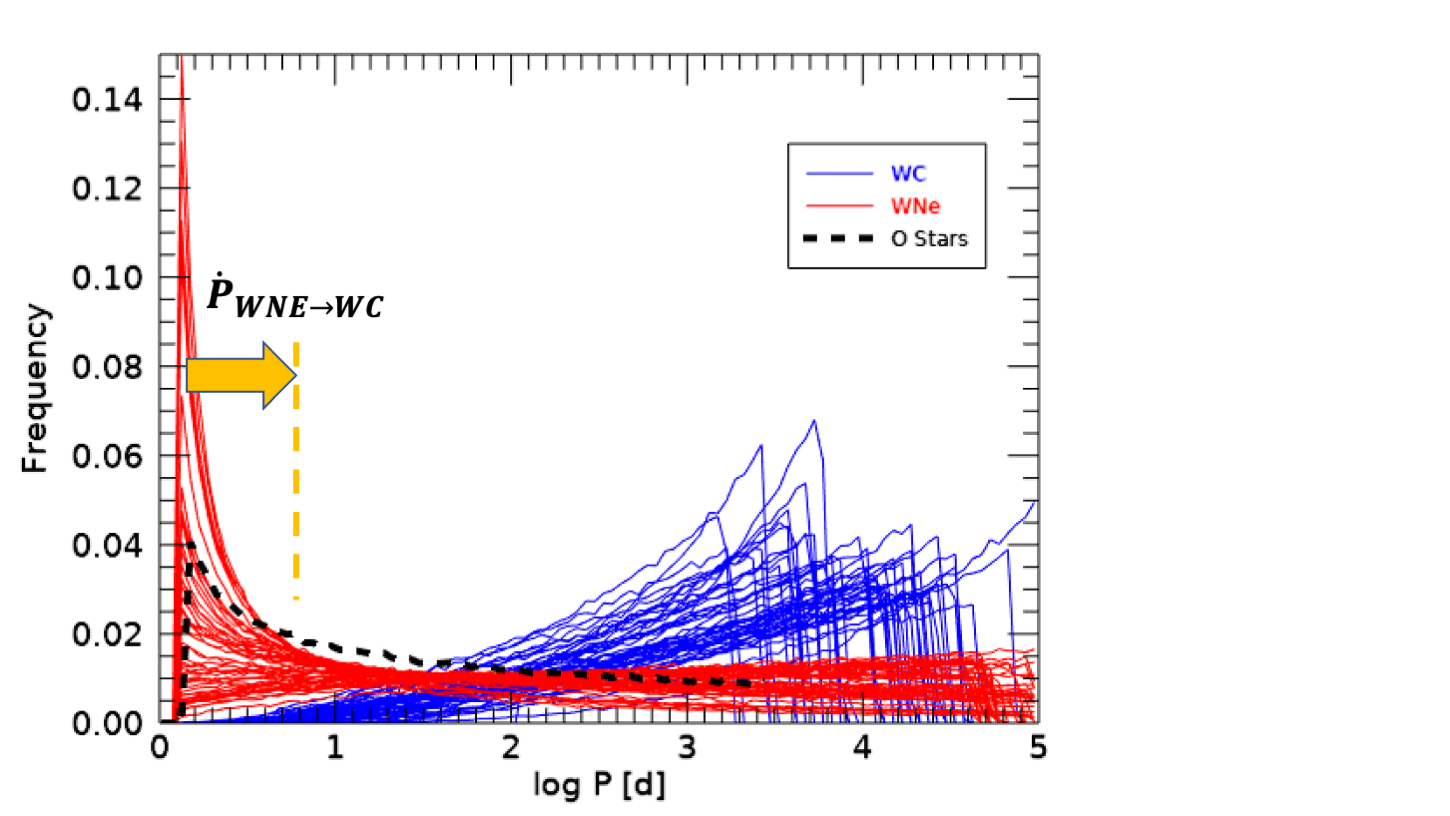} 
 \caption{Preliminary measured distributions of the WNE (red) and WC (blue) compared to that of O-type stars (black). Each curve follows $f_{\log P}\propto(\log P)^\pi$ and is characterised by a maximum cutoff period $P_\mathrm{max}$. The WNE and WC curves provide 50 realisations drawn from the $\pi$ and $P_\mathrm{max}$ posterior distributions of \cite{DSS22}. The yellow arrow indicates the expected widening of the orbits from the WNE to the WC phases  resulting from angular momentum loss due to stellar winds. See  \cite{DSS22} for final results.}
   \label{f:PdistWR}
\end{center}
\end{figure}

\subsection{classical Wolf-Rayet stars}

Classical Wolf-Rayet stars (cWRs) are (post)-Helium burning massive stars that have been stripped of their hydrogen envelope, revealing their inner core that has been enriched in nucleosynthesis products. Depending on the chemical compositions of their outer atmosphere, one distinguishes three main categories of cWRs: the nitrogen-, carbon-, and oxygen-rich classical Wolf-Rayet (WNs, WCs, and WOs, respectively).  Because of their luminosities, temperatures and small sizes, the radiation pressure on the atomic lines in the far-ultraviolet domain drives a very strong, optically thick stellar wind. Observationally, the wind signatures dominate their spectral appearance, with strong and often broad emission lines of He, C, N and O. 

In single stars, the stripping of the envelope is expected to be due to stellar winds, successively  revealing deeper and deeper layers, offering an evolutionary progression from WN to WC to, in a small number of cases, WO. In the Milky Way, one estimates a minimum initial mass of 20 to 30~M$_\odot$ (\cite{SGV20}) for a  single star to be able to strip itself to the cWR stage through stellar winds. In a close binary system, stripping might also occur through mass transfer once the initially more massive star fills its Roche lobe. In principle, binary interaction  decreases  the minimum initial mass needed to produce a Wolf-Rayet object, possibly contributing to populate a lower luminosity  region in the HRD than that occupied by wind-stripped cWRs (\cite{SGV20}).

Following the Conti scenario, early-type Wolf-Rayet stars of the Nitrogen sequence (WNE) will evolve into WC stars if they shed enough mass before exploding as SNe. 
\cite{DSS20} investigated  the multiplicity properties of two similar sized samples of WNE and WC stars. They found that the bias-corrected binary fractions of WC and WNE are respectively $>80$\%\ and $56\pm15$\%\, respectively. While error bars are large, the results remain intriguing. On the one hand,  how can WNE stars evolved into WC stars if their binary fraction is smaller? A possible explanation is that the RV sensitivity of WNE observations is limited by their wind-induced line profile variations, preventing a reliable probing at large orbital periods.  On the other hand, even the distributions of the orbital periods are puzzling. Looking at the Galactic Wolf-Rayet Star catalog (\cite{RC15}), \cite{DSS22} reported that most of the know WNE binaries have period of the order of 100 days or less. Most of the known WC binaries have periods of 1000 days or more (\cite{DSS20}). An initial attempt to estimate the true orbital period distribution of each sample from Dsilva et al. is displayed in Fig.~\ref{f:PdistWR}. We refer to \cite{DSS22} for updated results, but the conclusions remain. WNE and WC stars seem to have different period distributions, with the first one being heavily tweaked towards short period systems and the latter is flatter and favours long-period systems. Mass losses from the WNE to the WC phase  will inevitably take away angular momentum and broaden the orbit, but the impact on the period values is expected to be limited to about a factor of few, i.e., insufficient to reconcile the WNE and WC period distributions (Fig.~\ref{f:PdistWR}). These initial results require further investigations but illustrate the diagnostic power of the information contained in the multiplicity properties of these  evolved populations.

\section{Conclusions}

Stellar winds and binarity are ubiquitous phenomena among stars heavier than our Sun. Taken separately, they are both responsible for interesting physical processes, e.g., wind instabilities, clumping or mass transfer.
Both phenomena have strong impact on the mass of the stars, it is therefore not surprising that both are critical ingredients to properly compute the evolution of stars.
 
 In this short review, we outlined two avenues that jointly consider  stellar winds and binarity. On the one hand, we considered stars with strong outflows in binary systems, creating new physical processes. Their studies offer new diagnostics for stellar winds (e.g. enabling new ways to measure mass-loss rates) but also for the binary systems and their components. In the second avenue, we focused on the massive stars and discussed  evolutionary connections between various categories of young and evolved massive stars in the light of their  multiplicity properties. We show that the multiplicity properties of RSGs are compatible to first order with that of OB main-sequence stars under the caveat that binary interaction strongly impact most binaries with separation smaller than the typical RSG radius. This implies that a large fraction of systems should interact before any if the stars reach the RSG phase, suggesting the presence of a significant population of binary products. We report on new results by \cite{BSS20} suggesting that Be stars could potentially offer such a  population, by showing that binary interaction formation channel for Be star matches the lack of main-sequence + Be binaries.
 
We also looked at the properties of the evolved counterpart of the most massive stars, WRs and LBVs. We reported on recent results on LBVs and WRs which, though based on small samples, raised sufficiently intriguing results to deserve future investigations. While more data are certainly needed to offer full scale conclusions, this short review has shown that much insight is to be gained by considering both outflows and multiplicity as complementary diagnostics tools to better understand  stellar evolution in general.


\begin{thebibliography}{53}
\expandafter\ifx\csname natexlab\endcsname\relax\def\natexlab#1{#1}\fi

\bibitem[{Abt {et~al.} (1990)}]{AGL90}
Abt, H.~A., Gomez, A.~E., \& Levy, S.~G. 1990, The Astrophysical Journal
  Supplement Series, 74, 551

\bibitem[{Almeida {et~al.} (2017)}]{AST17a}
Almeida, L.~A., Sana, H., Taylor, W., {et~al.} 2017, Astronomy \& Astrophysics, 598, A84

\bibitem[{Banyard {et~al.} (2022)}]{BSM22}
Banyard, G., Sana, H., Mahy, L., {et~al.} 2022, Astronomy \& Astrophysics, 658, A69

\bibitem[{Bodensteiner {et~al.} (2020)}]{BSS20}
Bodensteiner, J., Shenar, T., \& Sana, H. 2020, Astronomy \&
  Astrophysics, 641, A42

\bibitem[{Bordier {et~al.} (2022)}]{BFS22}
Bordier, E., Frost, A.~J., Sana, H., {et~al.} 2022, Astronomy \&
  Astrophysics, in press (arXiv:2203.05036)
  
\bibitem[{Brott {et~al.} (2011)}]{BdC11a}
Brott, I.,  de Mink, S.~E., Cantiello, M., {et~al.} 2011, Astronomy \&
  Astrophysics, 513, A115

\bibitem[{Callingham {et~al.} (2019)}]{CTP19}
Callingham, J.~R., Tuthill, P.~G., Pope, B. J.~S., {et~al.} 2019, Nature
  Astronomy, 3, 82

\bibitem[{Clementel {et~al.} (2015{\natexlab{a}})}]{CMK15}
Clementel, N., Madura, T.~I., Kruip, C. J.~H., Paardekooper, J.~P., \& Gull,
  T.~R. 2015{\natexlab{b}}, Monthly Notices of the Royal Astronomical Society,
  447, 2445

\bibitem[{Clementel {et~al.} (2015{\natexlab{b}})}]{CMK15a}
Clementel, N., Madura, T.~I., Kruip, C. J.~H., \& Paardekooper, J.~P.
  2015{\natexlab{a}}, Monthly Notices of the Royal Astronomical Society, 450,
  1388

\bibitem[{Conti (1976)}]{C76a}
Conti, P. 1976, Mem. Soc. Roy. Sciences Li\`ege, IX, 129

\bibitem[{{de Koter} {et~al.} (1997)}]{dHH97}
{de Koter}, A., Heap, S.~R., \& Hubeny, I. 1997, The Astrophysical Journal,
  477, 792

\bibitem[{Decin {et~al.} (2020)}]{DMR20}
Decin, L., Montarg{\`e}s, M., Richards, A. M.~S., {et~al.} 2020, Science, 369, 1497

\bibitem[{Dorda \& Patrick (2021)}]{DP21}
Dorda, R. \& Patrick, L.~R. 2021, Monthly Notices of the Royal Astronomical
  Society, 502, 4890

\bibitem[{Dougherty {et~al.} (2005)}]{DBC05}
Dougherty, S.~M., Beasley, A.~J., Claussen, M.~J., Zauderer, B.~A., \&
  Bolingbroke, N.~J. 2005, The Astrophysical Journal, 623, 447

\bibitem[{Dsilva {et~al.} (2020)}]{DSS20}
Dsilva, K., Shenar, T., Sana, H., \& Marchant, P. 2020, Astronomy \&
  Astrophysics, 641, A26

\bibitem[{Dsilva {et~al.} (2022)}]{DSS22}
Dsilva, K., Shenar, T., Sana, H., \& Marchant, P. 2022, Astronomy \&
  Astrophysics, in press

\bibitem[{Duch{\^e}ne \& Kraus (2013)}]{DK13}
Duch{\^e}ne, G. \& Kraus, A. 2013, Ann. R. Astron. Astroph., 51, 269

\bibitem[{Dunstall {et~al.} (2015)}]{DDS15a}
Dunstall, P.~R., Dufton, P.~L., Sana, H., {et~al.} 2015, Astronomy \& Astrophysics, 580, A93

\bibitem[{Grellmann {et~al.} (2013)}]{GPR13}
Grellmann, R., Preibisch, T., Ratzka, T., {et~al.} 2013, Astronomy \& Astrophysics, 550, A82

\bibitem[{Humphreys {et~al.} (2016)}]{HWD16}
Humphreys, R.~M., Weis, K., Davidson, K., \& Gordon, M.~S. 2016, The
  Astrophysical Journal, 825, 64

\bibitem[{Johnstone {et~al.} (2015)}]{JZP15}
Johnstone, C.~P., Zhilkin, A., {Pilat-Lohinger}, E., {et~al.} 2015, Astronomy
  \& Astrophysics, Volume 577, A122

\bibitem[{Karl {et~al.} (2018)}]{KPE18}
Karl, M., Pfuhl, O., Eisenhauer, F., {et~al.} 2018, Astronomy \& Astrophysics,strik
  620, A116

\bibitem[{Kee {et~al.} (2021)}]{KSD21}
Kee, N.~D., Sundqvist, J.~O., Decin, L., de~Koter, A., \& Sana, H. 2021,
  Astronomy \& Astrophysics, 646, A180

\bibitem[{Lamberts {et~al.} (2012)}]{LDL12}
Lamberts, A., Dubus, G., Lesur, G., \& Fromang, S. 2012, Astronomy \&
  Astrophysics, 546, A60

\bibitem[{Lamers \& Levesque(2017)}]{LL17}
Lamers, H. \& Levesque, E. 2017, Understanding {{Stellar Evolution}} ({IOP
  Publishing Ltd})

\bibitem[{Lau {et~al.} (2020)}]{LHH20}
Lau, R.~M., Hankins, M.~J., Han, Y., {et~al.} 2020, The Astrophysical Journal,
  900, 190

\bibitem[{L{\"u}hrs (1997)}]{L97}
L{\"u}hrs, S. 1997, Publ. Astron. So. Pac., 109, 504

\bibitem[{Madura {et~al.} (2013)}]{MGO13}
Madura, T.~I., Gull, T.~R., Okazaki, A.~T., {et~al.} 2013, Monthly Notices of
  the Royal Astronomical Society, 436, 3820

\bibitem[{Mahy {et~al.} (2022)}]{MLH22}
Mahy, L., Lanthermann, C., Hutsem{\'e}kers, D., {et~al.} 2022, Astronomy \&
  Astrophysics, 657, A4

\bibitem[{Moe \& Di~Stefano (2017)}]{MD17}
Moe, M. \& Di~Stefano, R. 2017, The Astrophysical Journal Supplement Series,
  230, 15
  
\bibitem[{Neugent {et~al.} (2020)}]{NLM20}
Neugent, K.~F., Levesque, E.~M., Massey, P., Morrell, N.~I., \& Drout, M.~R.
  2020, The Astrophysical Journal, 900, 118

\bibitem[{Parkin \& Gosset (2011)}]{PG11}
Parkin, E.~R. \& Gosset, E. 2011, Astronomy and Astrophysics, 530, A119

\bibitem[{Patrick {et~al.} (2019)}]{PLB19a}
Patrick, L.~R., Lennon, D.~J., Britavskiy, N., {et~al.} 2019, Astronomy \&
  Astrophysics, 624, A129

\bibitem[{Patrick {et~al.} (2020)}]{PLE20a}
Patrick, L.~R., Lennon, D.~J., Evans, C.~J., {et~al.} 2020, Astronomy \&
  Astrophysics, 635, A29
  
\bibitem[{Pittard(2009)}]{P09}
Pittard, J.~M. 2009,  Monthly Notices of the Royal Astronomical Society, 396, 1743

\bibitem[{{Ram{\'i}rez-Tannus} {et~al.} (2021)}]{RBd21}
{Ram{\'i}rez-Tannus}, M.~C., Backs, F., {de Koter}, A., {et~al.} 2021,
  Astronomy \& Astrophysics, 645, L10

\bibitem[{{Ram{\'i}rez-Tannus} {et~al.} (2017)}]{RKK17}
{Ram{\'i}rez-Tannus}, M.~C., Kaper, L., de~Koter, A., {et~al.} 2017, Astronomy
  \& Astrophysics, 604, A78

\bibitem[{Rauw {et~al.} (2002)}]{RVS02}
Rauw, G., Vreux, J.-M., Stevens, I.~R., {et~al.} 2002, Astronomy \& Astrophysics, 388, 552

\bibitem[{Rosslowe \& Crowther (2015)}]{RC15}
Rosslowe, C.~K. \& Crowther, P.~A. 2015, Monthly Notices of the Royal
  Astronomical Society, 447, 2322

\bibitem[{Sana {et~al.} (2004)}]{SSG04}
Sana, H., Stevens, I.~R., Gosset, E., Rauw, G., \& Vreux, J.-M. 2004,
   Monthly Notices of the Royal Astronomical Society, 350, 809

\bibitem[{Sana {et~al.} (2012)}]{Sdd12}
Sana, H., {de Mink}, S.~E., {de Koter}, A., {et~al.} 2012, Science, 337, 444

\bibitem[{Sana {et~al.} (2013{\natexlab{a})}}]{SKM13}
Sana, H., {de Koter}, A., {de Mink}, S.~E., {et~al.} 2013,  Astronomy \& Astrophysics,
  550, A107

\bibitem[{Sana {et~al.} (2013{\natexlab{b})}}]{SvT13}
Sana, H., {van Boeckel}, T., Tramper, F., {et~al.} 2013,  Monthly Notices of the Royal Astronomical Society,
  432, L26

\bibitem[{Sana {et~al.} (2014)}]{SLL14}
Sana, H., Le~Bouquin, J.-B., Lacour, S., {et~al.} 2014, The Astrophysical Journal Supplement Series,
  215, 15

\bibitem[{Sana {et~al.} (2017)}]{SRK17}
Sana, H., {Ram{\'i}rez-Tannus}, M.~C., de~Koter, A., {et~al.} 2017, Astronomy
  \& Astrophysics, 599, L9
  
\bibitem[{Shenar {et~al.} (2020{\natexlab{a}})}]{SGV20}
Shenar, T., Gilkis, A., Vink, J.~S., Sana, H., \& Sander, A. A.~C.
  2020{\natexlab{a}}, Astronomy \& Astrophysics, 634, A79

\bibitem[{Shenar {et~al.} (2020{\natexlab{b}})}]{SSH20}
Shenar, T., Sablowski, D.~P., Hainich, R., {et~al.} 2020{\natexlab{b}},
  Astronomy \& Astrophysics, 641, C2

\bibitem[{Shenar {et~al.} (2021)}]{SSM21}
Shenar, T., Sana, H., Marchant, P., {et~al.} 2021, Astronomy \& Astrophysics,
  650, A147

\bibitem[{Smith (2019)}]{S19}
Smith, N. 2019, Monthly Notices of the Royal Astronomical Society, 489, 4378

\bibitem[{Smith \& Tombleson (2015)}]{ST15}
Smith, N. \& Tombleson, R. 2015, Monthly Notices of the Royal Astronomical
  Society, 447, 598

\bibitem[{Soulain {et~al.} (2018)}]{SML18}
Soulain, A., Millour, F., Lopez, B., {et~al.} 2018, Astronomy \& Astrophysics,
  618, A108

\bibitem[{Stevens {et~al.} (1992)}]{SBP92}
Stevens, I.~R., Blondin, J.~M., \& Pollock, A. M.~T. 1992, The Astrophysical Journal,
  386, 265

\bibitem[{Thomas {et~al.} (2021)}]{TRE21a}
Thomas, J.~D., Richardson, N.~D., Eldridge, J.~J., {et~al.} 2021, Monthly
  Notices of the Royal Astronomical Society, 504, 5221

\bibitem[{Thompson {et~al.} (2012)}]{TEM12}
Thompson, S.~E., Everett, M., Mullally, F., {et~al.} 2012, The Astrophysical
  Journal, 753, 86

\bibitem[{Tuthill {et~al.} (1999)}]{TMD99}
Tuthill, P.~G., Monnier, J.~D., \& Danchi, W.~C. 1999, Nature, 398, 487

\bibitem[{Tuthill {et~al.} (2008)}]{TML08}
Tuthill, P.~G., Monnier, J.~D., Lawrance, N., {et~al.} 2008, The Astrophysical
  Journal, 675, 698

\bibitem[{Usov (1992)}]{U92}
Usov, V.~V. 1992, The Astrophysical Journal, 389, 635

\bibitem[{Villase\~nor {et~al.} (2021)}]{VTE21}
Villase\~nor, J.~I., Taylor, W.~D., Evans, C.~J, {et~al.} 2021, Monthly Notices of the Royal Astronomical Society, 507,
  5348

\bibitem[{Walborn {et~al.} (2014)}]{WSS14}
Walborn, N.~R., Sana, H., {Sim{\'o}n-D{\'i}az}, S., {et~al.} 2014, Astronomy \& Astrophysics, 564,
  A40

\end{thebibliography}
\end{document}